\documentstyle[aps,prl]{revtex}
\input{psfig}
\draft

\begin{document}
\preprint{FERMILAB-Pub-98/178-E, EFI-98-21}

\title{ Search for the decay $K_L \rightarrow \pi^0 \nu \overline{\nu}$} 
\author{
J.~Adams$^{11}$,
A.~Alavi-Harati$^{12}$,
I.F.~Albuquerque$^{10}$,
T.~Alexopoulos$^{12}$,
M.~Arenton$^{11}$,
K.~Arisaka$^2$,
S.~Averitte$^{10}$,
A.R.~Barker$^5$,
L.~Bellantoni$^7$,
A.~Bellavance$^9$,
J.~Belz$^{10}$,
R.~Ben-David$^7$,
D.R.~Bergman$^{10}$,
E.~Blucher$^4$, 
G.J.~Bock$^7$,
C.~Bown$^4$, 
S.~Bright$^4$,
E.~Cheu$^1$,
S.~Childress$^7$,
R.~Coleman$^7$,
M.D.~Corcoran$^9$,
G.~Corti$^{11}$, 
B.~Cox$^{11}$,
M.B.~Crisler$^7$,
A.R.~Erwin$^{12}$,
S.~Field$^2$,
R.~Ford$^7$,
G.~Graham$^4$, 
J.~Graham$^4$,
K.~Hagan$^{11}$,
E.~Halkiadakis$^{10}$,
K.~Hanagaki$^8$, 
M.~Hazumi$^8$, 
S.~Hidaka$^8$,
V.~Jejer$^{11}$,
J.~Jennings$^2$,
D.A.~Jensen$^7$,
P.T.~Johnson$^7$,
R.~Kessler$^4$,
H.G.E.~Kobrak$^{3}$,
J.~LaDue$^5$,
A.~Lath$^{10}$,
A.~Ledovskoy$^{11}$,
A.P.~McManus$^{11}$,
P.~Mikelsons$^5$,
S.~Mochida$^8$,
E.~Monnier$^{4,*}$,
T.~Nakaya$^{7,\dagger}$,
U.~Nauenberg$^5$,
K.S.~Nelson$^{11}$,
H.~Nguyen$^7$,
V.~O'Dell$^7$, 
M.~Pang$^7$, 
R.~Pordes$^7$,
V.~Prasad$^4$, 
C.~Qiao$^4$, 
B.~Quinn$^4$,
E.J.~Ramberg$^7$, 
R.E.~Ray$^7$,
A.~Ronzhin$^7$,
A.~Roodman$^4$, 
M.~Sadamoto$^8$, 
S.~Schnetzer$^{10}$,
K.~Senyo$^8$, 
P.~Shanahan$^7$,
P.S.~Shawhan$^4$, 
W.~Slater$^2$,
N.~Solomey$^4$,
S.V.~Somalwar$^{10}$, 
R.L.~Stone$^{10}$, 
I.~Suzuki$^8$,
E.C.~Swallow$^{4,6}$,
R.A.~Swanson$^{3}$,
S.A.~Taegar$^1$,
R.J.~Tesarek$^{10}$, 
G.B.~Thomson$^{10}$,
P.A.~Toale$^5$,
A.~Tripathi$^2$,
R.~Tschirhart$^7$, 
Y.W.~Wah$^4$,
H.B.~White$^7$, 
J.~Whitmore$^7$,
B.~Winstein$^4$, 
R.~Winston$^4$, 
J.-Y.~Wu$^5$,
T.~Yamanaka$^8$,
E.D.~Zimmerman$^4$
}

\author{(KTeV Collaboration)\vspace{.5cm}}

\address{
$^1$ University of Arizona, Tucson, Arizona 85721 \\
$^2$ University of California at Los Angeles, Los Angeles, California 90095 \\
$^{3}$ University of California at San Diego, La Jolla, California 92093 \\
$^4$ The Enrico Fermi Institute, The University of Chicago, 
Chicago, Illinois 60637 \\
$^5$ University of Colorado, Boulder, Colorado 80309 \\
$^6$ Elmhurst College, Elmhurst, Illinois 60126 \\
$^7$ Fermi National Accelerator Laboratory, Batavia, Illinois 60510 \\
$^8$ Osaka University, Toyonaka, Osaka 560 Japan \\
$^9$ Rice University, Houston, Texas 77005 \\
$^{10}$ Rutgers University, Piscataway, New Jersey 08855 \\
$^{11}$ The Department of Physics and the Institute of Nuclear and 
Particle Physics, University of Virginia, 
Charlottesville, Virginia 22901 \\
$^{12}$ University of Wisconsin, Madison, Wisconsin 53706 \\
$^{*}$ On leave from C.P.P. Marseille/C.N.R.S., France \\
\normalsize
}
\maketitle

\vspace{-5.2in}
\begin{flushright}
EFI-98-21, FERMILAB-Pub-98/178-E\\
%FERMILAB-PUB-97-320-E
\end{flushright}
\vspace{5.0in}
\begin{center}(EFI-98-21, FERMILAB-Pub-98/178-E, Submitted to Phys. Rev. Lett.)\end{center}

\begin{abstract}
We report on a search for the rare decay $K_L \rightarrow \pi^0 \nu
\overline{\nu}$ in the KTeV experiment at Fermilab. We searched for
two-photon events whose kinematics were consistent with an isolated
$\pi^0$ coming from the decay  $K_L \rightarrow \pi^0 \nu
\overline{\nu}$. One candidate event was observed, which was
consistent with the expected level of background. 
An upper limit on the branching ratio was determined to be
$B(K_L \rightarrow \pi^0 \nu \overline{\nu}) < 1.6\times 10^{-6}$ at
the $90\%$ confidence level.  

\vspace*{0.1in}

\end{abstract}


%\newpage
%\narrowtext
\twocolumn

The rare decay $K_L \rightarrow \pi^0 \nu \overline{\nu}$ has
attracted considerable attention for the purpose of understanding the
phenomenology of CP violation because the decay rate results almost
entirely from a direct CP violating amplitude~\cite{kpinn1,kpinn2}.  
In the Wolfenstein parameterization of the CKM matrix~\cite{ckm,wolf},
the decay rate is proportional to the square of the parameter $\eta$, 
the value of which governs the scale of all CP violating
phenomena in the Standard Model. The reliability of the decay rate
calculation makes the decay $K_L \rightarrow \pi^0 \nu \overline{\nu}$
one of the best tools to determine the value of
$\eta$~\cite{kpinn3,kpinn4}. The predictions for the  branching ratio
are in the range of $(2.0-4.0)\times 10^{-11}$ using current knowledge
of Standard Model parameters, where the theoretical contribution to
the uncertainty is on the order of $1\%$~\cite{kpinn3}. Therefore,
observation of the decay near the predicted rate would provide
quantitative evidence in support of the Standard Model description of
CP violation, while observation outside the predicted rate would be
evidence of new physics~\cite{new}.

Although the decay is very clean theoretically, it is very challenging
experimentally. The experimental difficulty arises from searching for
a single isolated $\pi^0$ in the presence of the much more abundant
$\pi^0$'s from $K_L \rightarrow 2 \pi^0$ and $K_L \rightarrow 3
\pi^0$ decays. The most sensitive search conducted thus far used  
the Dalitz decay, $\pi^0 \rightarrow e^+ e^- \gamma$, to identify the
$\pi^0$, and set an upper limit of $5.8 \times 10^{-5} (90\% \mbox{ 
C.L.})$ on the branching ratio~\cite{e799}. 
While requiring a $\pi^0$~Dalitz decay provides a more unambiguous
experimental signature, this technique is at least two orders of
magnitude less sensitive than a direct search using $\pi^0 \rightarrow
\gamma \gamma$ due to the smaller branching ratio and lower
experimental acceptance.  A search using $\pi^0 \rightarrow \gamma
\gamma$ is the most viable method of reaching the predicted
sensitivity, and is the motivation for the experiment presented here.

\begin{figure}[hbt]
  \centerline{ \psfig{figure=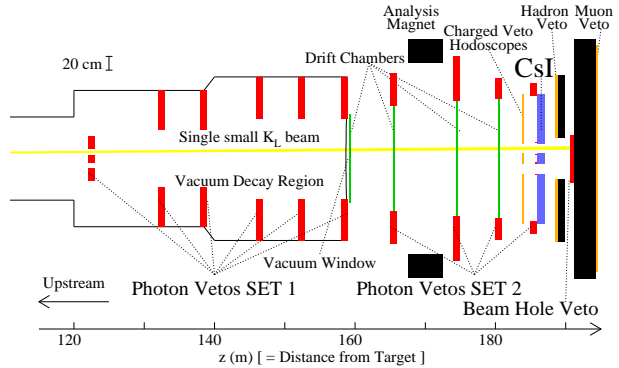,width=8.5cm} }
  \caption{Plan view of the KTeV detector as configured for this data
  set. \label{det}} 
\end{figure}

The first search for the decay $K_L \rightarrow \pi^0 \nu
\overline{\nu}$ using the $\pi^0 \rightarrow \gamma \gamma$ decay mode
was conducted in the KTeV experiment at Fermilab with data collected
in a one-day special run. 
A plan view of the KTeV detector is shown in Figure~\ref{det}. 
For this measurement, a single neutral beam was
produced by an 800~GeV proton beam that struck a $30$~cm long
beryllium-oxide target. The neutral beam was collimated to be
$0.065$~$\mu \rm str$ in solid angle 
($3.6 \times 3.6$ $\rm cm^2$ at 159~m downstream of the target), 
which was narrower than the
nominal KTeV beam size of $0.25$~$\mu \rm str$. The narrower beam
provides a tighter constraint on the  decay position in the
plane transverse to the beam direction, 
which improves our ability to measure the transverse momentum of the
$\pi^0$. The beam entered an evacuated decay volume which extended
from $90$~m to $159$~m downstream of the target. 
The downstream end of the volume was
closed by a vacuum window made of Kevlar and Mylar with a total
thickness of $0.002$ radiation lengths. 
The vacuum was maintained below $1.0
\times 10^{-4}$ torr.  The neutral beam was composed of neutrons and 
$K_L$ with the relative ratio of $2:1$. A small component of other
long-lived neutral particles, such as $\Lambda$, $\overline{\Lambda}$,
$\Xi^0$, and $\overline{\Xi}^0$ was also present. The beam flux of
neutral hadrons resulting from $3.5 \times 10^{12}$ protons per spill
on target was approximately $3 \times 10^7$.
The average kaon momentum was $70$ $\rm GeV/c$.
%--- Fig-1

The most crucial detector elements for this analysis are
a pure CsI electromagnetic calorimeter and a photon veto system.
The calorimeter is comprised of 3100 blocks in a $1.9$~m by $1.9$~m
square array, which is 27 radiation lengths deep. The energy 
resolution for electrons with an energy between 2 and 60~GeV in this
data set is approximately $1 \%$.  The calorimeter 
has two $15 \times 15$~$\rm cm^2$ square holes which allow for the passage
of two kaon beams used by the $\varepsilon^{'} / \varepsilon$ and other
rare kaon decay measurements made by the KTeV experiment~\cite{ktev}.    
The photon veto system is composed of eleven sets of lead/scintillator
sandwich detectors (the set-1, set-2 and the beam hole veto).  
The photon veto detector located the farthest upstream has two
beam holes and defines the upstream fiducial decay volume at
$z=122$~m.  This veto detector is followed by nine sets of photon veto
detectors which provide hermetic photon coverage up to an angle of
100~mr with respect to the beam direction. 
The last photon veto detector is the beam hole veto (BHV) which is
located in  the beam region and comprises part of the beam dump. 
The BHV is designed to identify photons and neutrons passing through
the beam holes of the calorimeter. The BHV is longitudinally segmented
into three sections which are each 10~radiation lengths
($0.42$~interaction lengths) deep.   
The two upstream sections are used to identify photons and measure
their energy, and the downstream section is used  to identify
neutrons.  For this analysis, the charged spectrometer, which consists
of four planar drift chambers and an analyzing magnet with a momentum
kick of 205~$\rm MeV/c$, is used to veto events with a charged
particle. Upstream of the calorimeter, there are 
charged veto hodoscopes used to veto events containing a 
charged particle. Downstream of the calorimeter, 
another set of hodoscopes (the hadron veto), 
preceded by 10 cm of lead, is used to veto events with a hadron 
that hits the calorimeter. Farther downstream, there are three 
meters of iron followed by a hodoscope (the muon veto), 
which is used to veto events with a muon.   

The trigger was designed to accept $K_L \rightarrow \pi^0 \nu
\overline{\nu}$ and $K_L \rightarrow 2 \pi^0$ decays. The latter decay
mode was used to measure the kaon 
flux of the experiment. The calorimeter was required to have an energy
deposit greater than 5~GeV. All of the hodoscopes were required to
have no activity. Events with an energy deposit above $500$~MeV in the
photon veto detectors were rejected. A hardware cluster counter 
processor~\cite{hcc} selected events with two or four isolated clusters
of energy in the calorimeter with at least one block in
each cluster having an energy greater than 1~GeV.

In the offline analysis,
candidate events are selected by requiring exactly two photons in the 
final state, each with energy greater than 1~GeV.
Events with more than two clusters of energy above $0.1$ GeV in the
CsI calorimeter are rejected.  Nearby photons with overlapping energy  
deposits in the calorimeter could reconstruct as a single cluster.
These events are rejected by requiring the transverse profile 
of each of the clusters
to be consistent with that of a single electromagnetic shower.  To
detect low energy photons which miss the calorimeter, an 
energy threshold of 70~MeV is imposed on the set-1 photon veto system 
located inside the evacuated volume (see Figure~\ref{det}). 
An energy threshold of 5~MeV is applied to the
set-2 photon veto system located downstream of the vacuum window.  
The lower threshold on the set-2 photon vetos
is imposed in order to detect the lower energy particles which may result 
from neutron interactions in the detector.
An energy threshold of 1~GeV is applied to each section of the beam
hole veto to eliminate events in which a photon or a neutron passing
through the CsI beam holes hits the detector. Events containing
hits in drift chambers are eliminated in order to remove backgrounds
containing a charged particle that misses the charged veto
hodoscopes.   

\begin{figure}
  \centerline{ \psfig{figure=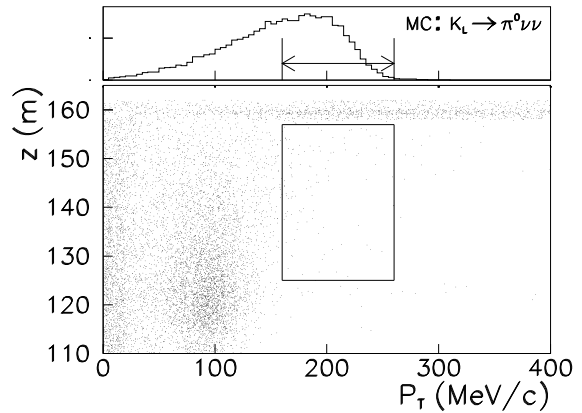,width=8.5cm}} 
  \caption{The top plot is the $P_T$ distribution of the $\pi^0$ from
  a Monte Carlo simulation of the decay $K_L \rightarrow \pi^0 \nu
  \overline{\nu}$. The bottom is the scatter plot of $P_T$ $\rm
  (MeV/c)$ versus the z decay vertex (m) for data after the initial
  event selection which requires exactly two photons in the final
  state. The box shows the signal region.  
\label{pt_zvtx}}  
\end{figure}

After the initial event selection, the z position of the decay vertex
of the event is calculated by requiring that the reconstructed
two-photon mass is consistent with a $\pi^0$. 
The z position is expressed as:
\begin{displaymath}
z  =  z_{CsI} - (r_{12}/m_{\pi^0})\sqrt{E_1 \cdot E_2},
\end{displaymath}
where $z_{CsI}$ is the z position of the calorimeter,
$r_{12}$ is the distance between the two photons in the calorimeter, 
$m_{\pi^0}$ is the $\pi^0$ mass and $E_i$ is the energy of the $i$-th
photon. The transverse momentum of the two photons relative to 
the beam direction, $P_T$, 
is also calculated by assuming that the decay occurred at the
transverse center of the beam.
An unbalanced $P_T$ distribution is a characteristic feature of the
decay $K_L \rightarrow \pi^0 \nu \overline{\nu}$, as shown in
Figure~\ref{pt_zvtx}. 
Figure~\ref{pt_zvtx} also shows a scatter plot of the $P_T$ spectrum
versus the z decay vertex for data. Neutral pions which are produced
by neutron interactions in the vacuum window at $\rm z = 159$~m can
be clearly seen. The events clustered around $P_T=100$~$\rm MeV/c$
and $\rm z = 120$~m are consistent with $\Lambda \rightarrow n \pi^0$ decays, where the
resulting $\pi^0$ can have a $P_T$ as high as $104$~$\rm MeV/c$. To
reduce these backgrounds, only those decays which have a reconstructed
z vertex between 125 and 157~m and $P_T$ between
160 and 260~$\rm MeV/c$ are selected. The $P_T$ cut at 260~$\rm MeV/c$
corresponds to the kinematic limit for $K_L \rightarrow \pi^0 \nu
\overline{\nu}$ decays, allowing for resolution effects. The
requirements on the z vertex and $P_T$ decrease the signal acceptance
by $57\%$; 177 out of 18,586 events pass these requirements. 
%--- Fig-2
%

Additional background results from $K_L \rightarrow 3 \pi^0$ events
which decay upstream of the photon veto detector at $\rm z = 122$~m. 
In these events, two photons from different $\pi^0$'s pass 
through the beam holes of this detector while the remaining photons 
go undetected. The decay vertex position of such events which are
misreconstructed as a $\pi^0$ can be shifted downstream into 
the fiducial region. These events can be identified by their 
large horizontal transverse momentum, $P_x$, because these
photons pass through the two horizontally separated beam holes 
of the detector. The ratio of $P_x/P_T$ is required to be less than  
$0.8$ in order to suppress this background.

Most of the $\Lambda$'s which decay in the detector are produced in the
target and are eliminated by the $P_T$ cut described
above.  A small fraction of the observed $\Lambda$ decays, however,
are the result of $\Xi^0 \rightarrow \Lambda \pi^0$ 
decays which yield $\pi^0$'s with a larger $P_T$ than those from
primary $\Lambda$'s.
Since the $\Lambda$'s have higher momentum than kaons on average,
the $\Lambda$ background is suppressed by requiring 
the total energy in the calorimeter to be less than 35 $\rm GeV$,
which results in a $30 \%$ loss in sensitivity for the 
decay $K_L \rightarrow \pi^0 \nu \overline{\nu}$.
The $\Lambda$ background can be further suppressed by eliminating
events with very asymmetric photon energies. The ratio of the 
maximum photon energy to the minimum photon energy is required to be
less than 6. 

\begin{figure}[hbt]
  \centerline{ \psfig{figure=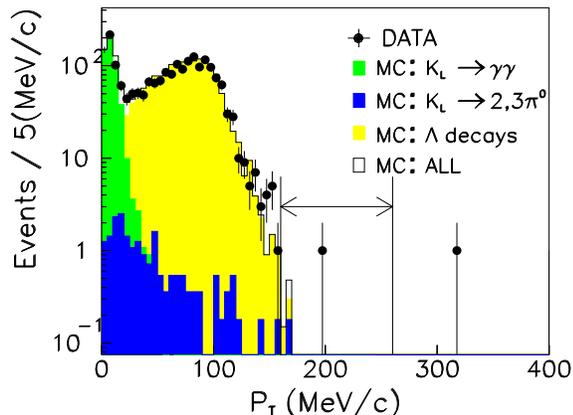,width=8.5cm} }
  \caption{ The $P_T$ distribution with all cuts except for the
  $P_T$ cut. The arrow shows the signal region.
  The dominant backgrounds from simulations of $K_L$ and $\Lambda$ 
  decays are overlaid on the data.
  $\Lambda$ events from the target and from $\Xi$ decays are 
  combined. There are no events beyond the $P_T$ range in the plot.  
  \label{pt_final}}
\end{figure}

Figure~\ref{pt_final} shows the $P_T$ distribution for events passing  
all cuts except for the $P_T$ cut. One event remains in the signal
region, and another event appears in the region above the
$P_T$ window. As discussed below, these two remaining events are
consistent with background events from neutron interactions. 
Neutrons which interact with material such as
the vacuum window or the drift chambers can produce one or 
more $\pi^0$'s or $\eta$'s. 
Interactions which produce $\eta$'s are potential background sources 
since the z~vertex is miscalculated due to the $\pi^0$ mass 
assignment. Interactions which produce multiple $\pi^0$'s are
additional background sources since the z~vertex can be
misreconstructed inside the fiducial volume when only two photons from
different $\pi^0$'s are detected.
%--- Fig-3

The background from neutron interactions that produce either
$\eta$'s or multiple $\pi^0$'s was estimated by removing
the requirement on the BHV and then determining how many of these
events have an undetected neutron. Even when the neutron is not
detected, 
one of the photons from multiple $\pi^0$'s can hit the BHV, and  
must be accounted for in the background estimate.
Using a clean sample of $\Lambda \rightarrow n \pi^0$ decays, the BHV
efficiency for a neutron was measured to be $44 \%$ and $80 \%$ in the 
upstream and downstream sections, respectively.
When the BHV cut is removed, 37 events remain. 
The number of events detected in the downstream section (31)
is consistent with the expectation from neutrons, but the
number of events detected in the upstream section (26) is $60 \%$ 
larger than the expectation from neutrons. The excess in 
the upstream section is assumed to be due to an extra photon
hitting the BHV in $(47 \pm 13) \%$ of the events.
Using the neutron detection efficiency in all sections of the BHV ($82
\%$) and the probability that an extra photon hits the BHV, the
background estimate is $37 \times (1-0.82) \times [1-(0.47\pm0.13)]
= 3.5 \pm 0.9$~events. Therefore, we conclude that the one observed
event in the signal region is consistent with background from neutron 
interactions. The background from neutron interactions also accounts
for the event with $P_T$ above the signal region.

In addition to the neutron interaction background, background from
$K_L$, $\Lambda$, and $\Xi^0$ decays are estimated from Monte Carlo  
simulations with 7 times the statistics of the data.
The $P_T$ distribution below 160 $\rm MeV/c$ is well
reproduced by the simulation, as shown in Figure~\ref{pt_final}.  
The background level from $K_L$ decays is estimated to be
$0.2^{+0.4}_{-0.1}$ events in the signal region.
The background level from $\Lambda$ and $\Xi^0$ decays is estimated to 
be $0.4$ events which are already included in the estimate of the  
neutron interaction background due to the existence of a neutron in
the $\Lambda$ decay. 

To determine the number of kaon decays in the fiducial volume of the
detector, we select a sample of $K_L \rightarrow 2 \pi^0$ decays
using selection criteria similar to those used for the $K_L
\rightarrow \pi^0 \nu \overline{\nu}$ decay. The sample consists of
4326 $K_L \rightarrow 2 \pi^0$ events with an estimated background of
44 events. From a Monte Carlo simulation, the detector acceptances 
for the decays $K_L \rightarrow \pi^0 \nu \overline{\nu}$ and $K_L
\rightarrow 2 \pi^0$ are determined  to be $3.6 \%$ and $6.7\%$,
respectively, for kaon energies between 10 and 230~GeV within the
decay region between 120 and 160~m. 
The total systematic error is determined to be  $6.0\%$, including
uncertainties in the acceptance calculation due to possible trigger 
bias, energy measurement and 
the treatment of accidental activity in the detector.

Using $B(K_L \rightarrow 2 \pi^0) = (9.36 \pm 0.20) \times
10^{-4}$~\cite{pdg},  the number of kaon decays for the experiment is  
measured to be $(6.8 \pm 0.2)\times 10^7$. 
The single event sensitivity for the decay $K_L \rightarrow \pi^0 \nu
\overline{\nu}$ is $\rm [4.04 \pm 0.06(stat.) \pm 0.24(sys.)] \times 
10^{-7}$. Conservatively assuming that the last remaining event is
signal, the upper limit on the branching ratio is determined to be
$B(K_L \rightarrow \pi^0 \nu \overline{\nu}) < 1.6 \times 10^{-6}$ at
the $90 \%$ confidence level, where both statistical and systematic
errors are included~\cite{sys}.  The sensitivity of this search
represents a factor of $36$ improvement relative to the best previous
limit~\cite{e799}. Further improvement using this technique will
require a significant reduction in the amount of material in the beam.  

We gratefully acknowledge the support and effort of the Fermilab
staff and the technical staffs of the participating institutions for
their vital contributions.  This work was supported in part by the U.S. 
Department of Energy, The National Science Foundation and The Ministry of
Education and Science of Japan.  In addition, A.R.B., E.B. and S.V.S. 
acknowledge support from the NYI program of the NSF; A.R.B. and E.B. from 
the Alfred P. Sloan Foundation; E.B. from the OJI program of the DOE; 
K.H., T.N. and M.S. from the Japan Society for the Promotion of
Science.  P.S.S. acknowledges receipt of a Grainger Fellowship.

\end{document}